\documentclass{aa}  
\usepackage[colorlinks=true, linkcolor=red, urlcolor=cyan, citecolor=blue]{hyperref}
\usepackage{graphicx}
\usepackage{txfonts}
\usepackage{multirow}
\usepackage{pdflscape}
\usepackage{rotate}
\usepackage{color}
\usepackage{adjustbox}
\usepackage{lscape}
\usepackage{amsmath}
\usepackage[dvipsnames,svgnames,x11names]{xcolor}
\graphicspath{.}

\begin{document} 

\title{The eccentric behaviour of windy binary stars}

\author{M. I. Saladino \inst{1, 2}
             \and
            O. R. Pols\inst{1}
            }

\institute{Department of Astrophysics/IMAPP, Radboud University, P.O. Box 9010, 6500 GL Nijmegen, The Netherlands\\
              \email{[m.saladino; o.pols]@astro.ru.nl}
               \and
               Leiden Observatory, Leiden University, PO Box 9513, 2300, RA, Leiden, The Netherlands
               }

\date{Received April 4, 2019; accepted June 4, 2019}
 
\abstract{
Carbon-enhanced metal-poor stars, CH stars, barium stars and extrinsic S stars, among other classes of chemically peculiar stars, are thought to be the products of the interaction of low- and intermediate-mass binaries which occurred when the most evolved star was in the asymptotic giant branch (AGB) phase.
Binary evolution models predict that because of the large sizes of AGB stars, if the initial orbital periods of such systems are shorter than a few thousand days, their orbits should have circularised due to tidal effects.
However, observations of the progeny of AGB binary stars show that many of these objects have substantial eccentricities, up to $e\approx 0.9$.
In this work we explore the impact of wind mass transfer on the orbital parameters of AGB binary stars by performing numerical simulations in which the AGB wind is modelled using a hydrodynamical code and the dynamics of the stars is evolved using an N-body code.
We find that in most models the effect of wind mass transfer will contribute to the circularisation of the orbit, but on longer timescales than tidal circularisation if $e\lesssim 0.4$. 
We also find that for relatively low initial wind velocities and pseudo-synchronisation of the donor star, a structure resembling wind Roche-lobe overflow is observed as the stars approach periastron. 
In this case, the interaction between the gas and the star is stronger than when the initial wind velocity is high and the orbit shrinks while the eccentricity decreases. 
In one of our models wind interaction is found to pump the eccentricity of the orbit on a similar timescale as tidal circularisation.
However, since the orbit of this model is shrinking tidal effects will become stronger during the evolution of the system.
Although our study is based on a small sample of models, it offers some insight into the orbital evolution of eccentric binary stars interacting via winds. 
A larger grid of numerical models for different binary parameters is needed to test if a regime exists where hydrodynamical eccentricity pumping can effectively counteract tidal circularisation, and if this can explain the puzzling eccentricities of the descendants of AGB binaries. 
}


\maketitle

\section{Introduction} \label{p3:sec:introd}

A wide variety of objects are thought to result from interaction in asymptotic giant branch (AGB) binary systems. 
These include barium stars and CH stars \citep{keenan1942}, extrinsic S stars \citep{Smith+Lambert1988}, carbon-enhanced metal poor (CEMP) stars \citep{Beers+Christlieb2005}, and binary post-AGB stars \citep{vanWinckel2003a}.
Observations of these objects show that they have large eccentricities (up to $e \approx 0.9$) for relatively short orbital periods \citep[between 100-1000 days;][]{jorissen+1998, jorissen2, Hansen+2016, vanderSwaelmen+2017, Oomen+2018}. 
However, because of the large sizes of AGB stars, binary evolution models predict that such systems should have circularised due to tidal forces if their orbital periods were initially shorter than a few thousand days \citep{pols, izzard}.
Therefore, a mechanism that counteracts tidal interaction or that enhances the eccentricity of the binary after tidal circularisation is needed to explain the observed orbits of these systems. 

Several mechanisms that can pump the eccentricity during the evolution of the binary system have been proposed, such as the interaction of the binary with a circumbinary disk \citep{Artymowicz+1991, Artymowicz+Lubow1994, Dermine+2013}, phase-dependent mass loss \citep[e.g.][]{Soker2000, Bonavic+2008} or grazing envelope evolution \citep{Soker2015}.
Eccentricity pumping due to interaction with a circumbinary disk is given some observational support by the detected presence of such disks in binary post-AGB stars \citep{deRuyter+2006}. 
However, \cite{Rafikov2016} argues that in order for this mechanism to efficiently increase the eccentricity of the binary, the circumbinary disk must be very massive and long-lived compared to the inferred estimates for such parameters.  
\cite{Vos+2015} test different eccentricity pumping mechanisms, such as phase-dependent mass loss and interaction with a circumbinary disk in an attempt to explain the eccentricities of hot subdwarf binaries. 
However, they find that these proposed mechanisms are insufficient to reproduce their observed eccentricities.
In a recent study, \cite{Kashi+Soker2018} show that grazing envelope evolution could efficiently counteract tidal circularisation, but they only consider a single set of binary parameters.

In addition to these proposed eccentricity pumping mechanisms, detailed analytical studies of the orbital evolution of eccentric binary systems have been performed by \cite{Sepinsky+2007b}, \cite{Sepinsky+2009} \cite{Eggleton2006}, and \cite{Dosopoulou+Kalogera2016}. 
For instance, \cite{Eggleton2006} and \cite{Dosopoulou+Kalogera2016} derive the secular evolution of the semi-major axis and the eccentricity of an eccentric binary when interaction occurs via fast isotropic winds. 
However, several hydrodynamical studies have shown that wind interaction in AGB binaries can be quite different from the isotropic-wind mode \citep[e.g.][]{theuns1, shazrene1, Saladino+2018a, rochester}.
Therefore, in order to understand how wind mass transfer interaction in eccentric AGB binary systems impacts the orbital evolution of the system hydrodynamical simulations are needed.

Most of the current hydrodynamical studies of interacting binary systems have been performed for systems in circular orbits \citep[e.g.][]{theuns1, theuns2, val-borro, shazrene, Liu+2017, Chen+2017, Saladino+2018a}, while only a  handful of studies have investigated hydrodynamical models for eccentric binary stars \citep[e.g.][]{Church+2009, shazrene_thesis, Lajoie+Sills2011, vdHelm+2016, Kim+2017}. 
However, with the exception of \cite{vdHelm+2016}, most of the studies on eccentric binaries have focussed on understanding the mass transfer process, while little attention has been devoted to the effect of mass transfer on the orbital evolution of the binary. 
The complexity in performing such studies arises from the fact that in order to derive the change in the orbital parameters of the binary, the changes in both the orbital angular momentum and the orbital energy need to be known.
Additionally, in order to determine the change in the orbital angular momentum, the angular-momentum loss from the system as well as the mass-accretion efficiency onto the companion star are needed. 
Such parameters can be estimated from hydrodynamical simulations \citep[as has been done for circular orbits, see e.g.][]{theuns2, shazrene, rochester, Saladino+2018b}.
However, in order to study the change in the semi-major axis and eccentricity simultaneously, numerical models in which the dynamics of the stars is modelled in detail are needed because they permit to estimate the change in the orbital energy as the binary interacts.
Furthermore, the hydrodynamical models by \cite{Kim+2017} show that in the case of eccentric binary stars interacting via winds, the morphology of the outflow can differ considerably from the non-eccentric case. 
As shown in \citet[][hereafter Paper I]{Saladino+2018a} and \citet[][hereafter Paper II]{Saladino+2018b}, the evolution of the orbital parameters is strongly influenced by the morphology of the outflow.
In addition, in Paper II we show that within the numerical uncertainties we can measure the change in the semi-major axis dynamically from numerical simulations.

In order to understand if the puzzling eccentricities of the descendants of AGB binary systems can be explained by an episode of wind mass transfer, in this paper we perform an exploratory numerical study of low- and intermediate-mass eccentric binaries interacting via AGB winds.
In our simulations we couple a hydrodynamical code with a gravitational code to follow the evolution of the orbit.
This allows us to measure simultaneously not only the amount of angular momentum-loss and mass-accretion efficiency, but also the change in the semi-major axis and eccentricity of the system.

\section{Method}\label{p3:sec:method}

\begin{sidewaystable*}
\caption{Parameters of the models}
\centering
\label{p3:table:setup}
\begin{adjustbox}{width=\textwidth}
\begin{tabular}{c c c c c c c c c c c c c c c c c c c c}
\hline\hline
Name	&	$M_\mathrm{d}$	&	$M_\mathrm{a}$	&	$q$	&	$P$	&	$a$	&	$e$	&		$J_\mathrm{orb}$		&	$v_{pe}$	&	$v_{ap}$	&	$v_\infty$	&	$v_\infty/v_{pe}$	&	$v_\infty/v_{ap}$	&	$\Omega_\mathrm{spin}$		&	$R_\mathrm{L,1}|_{pe}$	&	$R_\mathrm{d}$	&	$R_\mathrm{sink}$	&	$T_\mathrm{eff}$	&	$\dot{M}_\mathrm{d}$	&		$m_\mathrm{g}$	\\
-	&	M$_\odot$	&	M$_\odot$	&	-	&	yr.	&	AU	&	-	&		m$^2$ kg s$^{-1}$		&	km s$^{-1}$	&	km s$^{-1}$	&	km s$^{-1}$	&	-	&	-	&	s$^{-1}$		&	R$_\odot$	&	R$_\odot$	&	R$_\odot$	&	K	&	M$_\odot$ yr$^{-1}$	&		M$_\odot$	\\ \hline
Q2e0	&	1.2	&	0.6	&	2	&	5.96	&	4.00	&	0	&	$	9.51	\times 10^{45}$	&	19.98	&	19.98	&	15.1	&	0.76	&	0.76	&	3.34	$\times 10^{-8}$	&	378.6	&	330	&	27.6	&	3240	&	$1.5 \times 10^{-5}$	&	$	9.6 \times 10^{-10}$	\\
Q2e02	&	1.2	&	0.6	&	2	&	8.33	&	5.00	&	0.2	&	$	1.04	\times 10^{46}$	&	21.89	&	14.59	&	15.1	&	0.69	&	1.03	&	3.66	$\times 10^{-8}$	&	376.3	&	330	&	27.2	&	3240	&	$1.5 \times 10^{-5}$	&	$	1.2 \times 10^{-9}$	\\
Q2e04	&	1.2	&	0.6	&	2	&	12.83	&	6.66	&	0.4	&	$	1.13	\times 10^{46}$	&	23.64	&	10.13	&	15.1	&	0.64	&	1.49	&	3.95	$\times 10^{-8}$	&	371.7	&	330	&	26.9	&	3240	&	$1.5 \times 10^{-5}$	&	$	1.6 \times 10^{-9}$	\\
Q2e06	&	1.2	&	0.6	&	2	&	23.57	&	10.00	&	0.6	&	$	1.20	\times 10^{46}$	&	25.28	&	6.32	&	15.1	&	0.60	&	2.39	&	4.22	$\times 10^{-8}$	&	365.9	&	330	&	26.8	&	3240	&	$1.5 \times 10^{-5}$	&	$	2.4 \times 10^{-9}$	\\
Q2e08	&	1.2	&	0.6	&	2	&	66.66	&	20.00	&	0.8	&	$	1.28	\times 10^{46}$	&	26.81	&	2.98	&	15.1	&	0.56	&	5.07	&	4.48	$\times 10^{-8}$	&	362.5	&	330	&	138.9	&	3240	&	$1.5 \times 10^{-5}$	&	$	4.8 \times 10^{-9}$	\\
Q2e06v1	&	1.2	&	0.6	&	2	&	23.57	&	10.00	&	0.6	&	$	1.20	\times 10^{46}$	&	25.28	&	6.32	&	6.0		&	0.24	&	0.95	&	4.22	$\times 10^{-8}$	&	365.9	&	330	&	26.8	&	3240	&	$1.5 \times 10^{-5}$	&	$	2.4 \times 10^{-9}$	\\
MMe05	&	3.0	&	1.5	&	2	&	5.27	&	5.00	&	0.5	&	$	3.64	\times 10^{46}$	&	48.95	&	16.32	&	14.2 &	0.31	&	0.93	&	0		&	231.3	&	200	&	16.7	&	2430	&	$1.5 \times 10^{-5}$	&	$	1.6 \times 10^{-9}$	\\
\hline \hline
\end{tabular}
\end{adjustbox}
\tablefoot{
The first column corresponds to the name of the model.
$M_\mathrm{d}$ and $M_\mathrm{a}$ are the masses of the donor and accretor star, and $q$ is the mass ratio.
$P$, $a$ and $e$ are the orbital period, semi-major axis and eccentricity. 
$J_\mathrm{orb}$ is the initial orbital angular momentum of the binary.
$v_{pe}$ and $v_{ap}$ are the orbital velocities at periastron and apastron, respectively, and $v_\infty$ is the terminal velocity of the wind.
$\Omega_\mathrm{spin}$ is the spin angular velocity of the donor star.
$R_\mathrm{L,1}|_{pe}$ is the Roche-lobe radius of the primary star at periastron. 
$R_\mathrm{d}$ is the radius of the donor star and
$R_\mathrm{sink}$ is the radius of the sink particle representing the companion star. 
$T_\mathrm{eff}$ is the effective temperature of the primary star which also determines the initial temperature of the wind particles. 
$\dot{M}_\mathrm{d}$ is the mass-loss rate of the donor star and
$m_\mathrm{g}$ is the SPH particle mass. 
}
\end{sidewaystable*}

\begin{figure*}
\centering
\includegraphics[width=0.97\hsize]{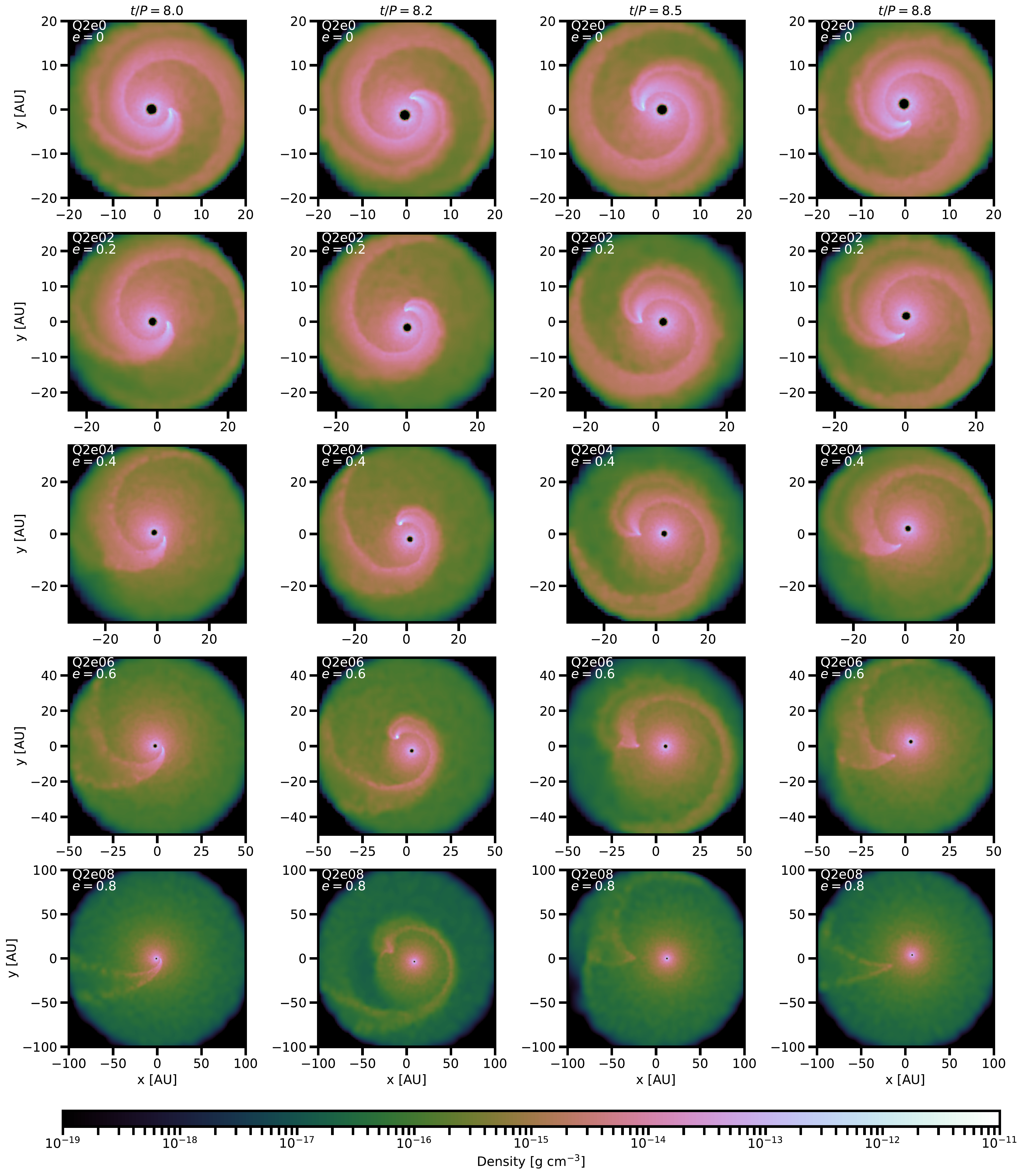}
\caption{
Gas density on the orbital plane for different orbital phases for models Q2e0 to Q2e08 during the ninth simulated orbit. 
The eccentricity increases from top to bottom. 
Time is indicated along the top, relative to the orbital period.
The first column corresponds to the phase where the stars are at periastron, the second column to the phase $t/P = 8.2$, the third column to the time when stars are at apastron and the fourth column correspond to $t/P = 8.8$.
}
\label{p3:fig:density1}
\end{figure*}

The numerical method {employed in this paper} is similar to that used in Paper I and Paper II.
Here we briefly describe the set-up of the simulations. 
Using the \textsc{amuse}\footnote{\url{http://www.amusecode.org/}} framework  \citep{amuse1, amuse2, amuse3} we couple the smoothed-particle hydrodynamics (SPH) code \textsc{fi} \citep{fi1, fi2, fi3} with the N-body code \textsc{huayno} \citep{huayno} using the \textsc{bridge} module \citep{bridge}.
The SPH code is used to model the gas dynamics of the wind while the N-body code is used to compute the dynamics of the stars. 
Our models also include the prescription for cooling or heating of the gas described in Paper II. 

The stars are modelled as point masses and the wind particles are created using the \textsc{stellar\_wind.py} module \citep{edwin} available in \textsc{amuse}. 
Wind particles are injected with initial velocities $v_\mathrm{init} = 12$ km s$^{-1}$ or $v_\mathrm{init} = 1$ km s$^{-1}$ at a spherical surface with the radius of the donor star, $R_\mathrm{d}$, and their initial temperature is equal to the effective temperature of the star ($T_\mathrm{eff}$). 
Similar to Paper II, the equation of motion of the wind in all models contains a term that exactly balances the gravity of the donor star, thus wind particles feel an extra acceleration due to gas pressure which drives them to an average terminal velocity $v_\infty \approx 15$ km s$^{-1}$ or $v_\infty \approx 6$ km s$^{-1}$, depending on the initial wind velocity.
These velocities correspond to the typical terminal velocities for AGB stars \citep{susanne}. 

To allow for comparison with our previous work, the stellar parameters for most of our models correspond to those described in Paper II for a mass ratio $q=M_\mathrm{d}/M_\mathrm{a} = 2$. 
These models have a metallicity of $Z = 10^{-4}$. 
Only the stellar parameters of model MMe05 are similar to those used in Paper I.
In this model the stars are more massive, the radius and effective temperature of the donor star are smaller and the donor metallicity is solar. 
In all models the donor star loses mass at a constant rate $\dot{M}_\mathrm{d} = 1.5 \times 10^{-5}$ M$_\odot$ yr$^{-1}$ (see Table \ref{p3:table:setup} for an overview of all parameters used). 
Each model is run for 10 orbital periods.

In Paper II, we find that the strength of interaction between the companion star and the wind depends on the wind-to-orbit velocity ratio, $v_\infty/v_\mathrm{orb}$. 
For large $v_\infty/v_\mathrm{orb}$ little interaction occurs and the outflow approximates the spherically symmetric wind mode, 
while the strongest interaction between the wind and the binary occurs for small $v_\infty/v_\mathrm{orb}$ and hence small separations. 
However, for eccentric binaries the relative orbital velocity and separation of the stars are time-variable.
Therefore, in order to guarantee a strong interaction at the point of closest approach, while ensuring that the donor star is within its Roche lobe at periastron, 
we set the semi-major axis of our models by keeping the periastron distance constant, $a_{pe} = 4$ AU, for eccentricities between 0 to 0.8.
We compute the Roche lobe radius at periastron using the equation given by \cite{Sepinsky+2007a} for a binary system in an eccentric orbit.

For a binary with the characteristics of our models, tidal effects are likely to pseudo-synchronise the donor star, i.e. the angular velocity of the star is assumed to be equal to the angular velocity of the binary at periastron. 
To this end, in a similar fashion to Paper II, we add a tangential velocity, $\mathbf{v}_\mathrm{T} = \mathbf{\Omega}_{\mathrm{orb}, pe} \times \mathbf{r}_\mathrm{g, d}$, to wind particles as we inject them at the surface of the star, where 
\begin{equation}\label{p3:eq:omegaper}
\Omega_{\mathrm{orb} , pe} = \frac{2\pi}{P} \frac{(1+e)^{1/2}}{(1-e)^{3/2}},
\end{equation}
is the orbital angular velocity at periastron, $P$ is the binary orbital period and $\mathbf{r}_\mathrm{g, d}$ is the position of the gas particles with respect to the centre of mass of the donor star. Only in model MMe05 the donor is non-rotating.

Similar to Paper I and Paper II, we model the companion star as a sink particle with constant radius. 
In the majority of the models the sink radius is equal to $0.1 R_{\mathrm{L,2}|pe}$, where $R_{\mathrm{L,2}|pe}$ is the Roche lobe radius  of the companion star at periastron. 
Only for model Q2e08, with $e = 0.8$, we set the sink radius equal to $0.5 R_{\mathrm{L,2}|pe}$.
The latter setup is chosen to prevent numerical artefacts in the simulation because the resolution of our models is low and for $e = 0.8$ the typical smoothing length in the vicinity of the companion star is much larger than a sink with radius $0.1 R_{\mathrm{L,2}|pe}$. 

To optimise the numerical computation, we choose the typical smoothing length of the particles to be proportional to the semi-major axis of the binary (see Paper II). 
In Table \ref{p3:table:setup} we show the corresponding masses of the gas particles. 
In addition, to minimise computational time we remove particles once they cross a boundary of $5a$. 
The values for the artificial viscosity parameters are set as $\alpha_\mathrm{SPH} = 0.5$ and $\beta_\mathrm{SPH} =1$.

\section{Results}\label{p3:sec:results}


\subsection{Morphology}\label{p3:sec:morphology}

\begin{figure*}
\centering
\includegraphics[width=0.97\hsize]{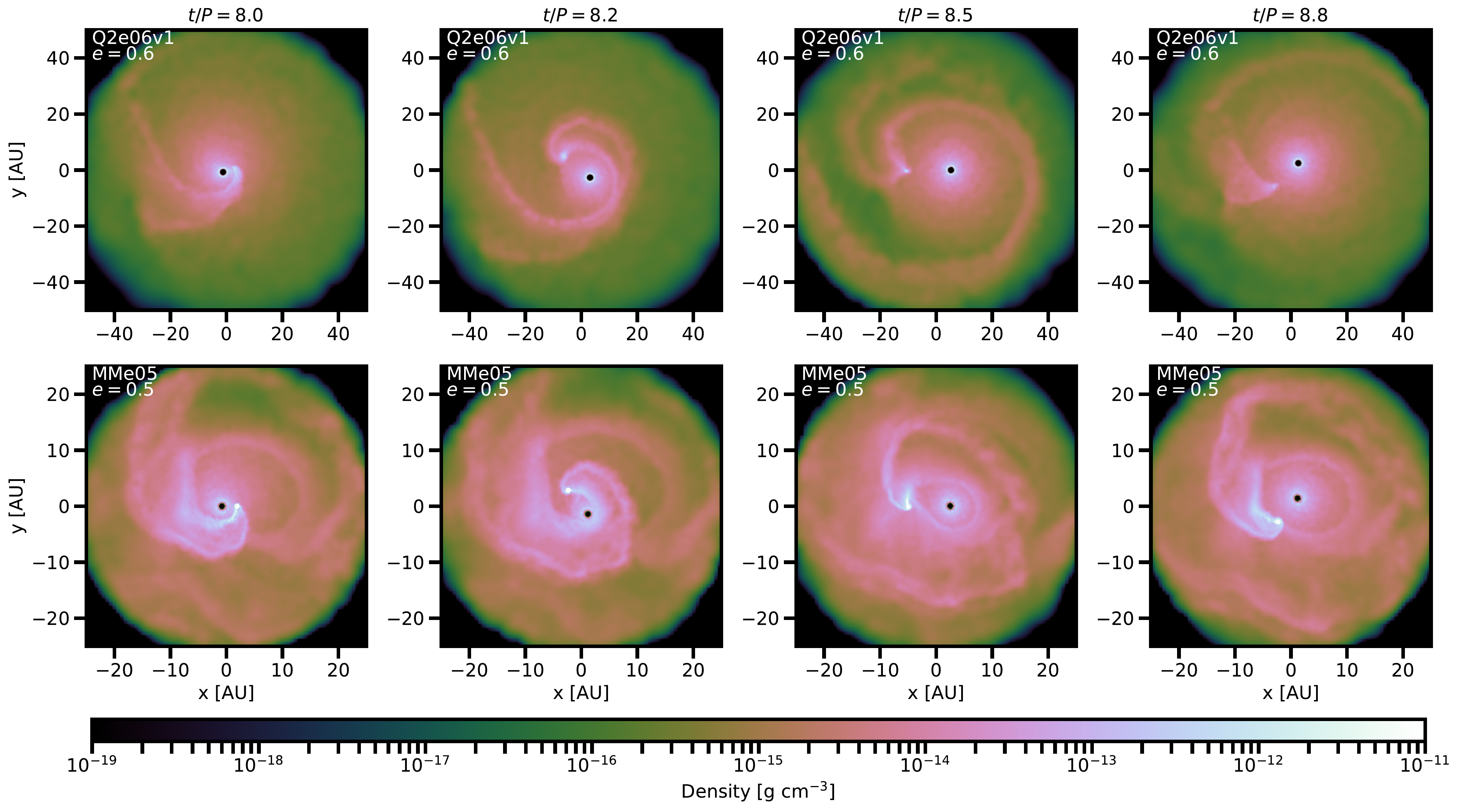}
\caption{Similar to Fig. \ref{p3:fig:density1}, but for models Q2e06v1 (top) and MMe05 (bottom).}
\label{p3:fig:density_q2_v1e5}
\end{figure*}

In the following we describe the observed morphologies of the outflow for the eccentric models and we compare them with the outflows observed in their circular counterparts. 
We note that although we only describe the behaviour of the outflow for one orbit, the same structures repeat over the evolution of the systems.
 
Figure \ref{p3:fig:density1} shows the density in the orbital plane at four orbital phases (left to right) for models Q2e0 to Q2e08.
The eccentricity of the models increases from top to bottom.  
For small eccentricities ($e = 0.2$ and $e = 0.4$) the geometry of the outflow looks very similar to the circular binary case with two spiral arms tightly wound around the binary delimiting the accretion wake behind the companion star.
Similar to the circular case, the inner spiral arm is denser than the outer one due to its proximity to the AGB donor star. 
In addition, the opening angle of the accretion wake varies as a function of the orbital phase and as a function of the eccentricity. 

As the eccentricity increases ($e = 0.6$ and $e = 0.8$) a gradual change in the morphology of the outflow is observed.
The accretion wake behind the companion star which for small eccentricities was wrapped around the binary in the form of spiral arms becomes a disrupted ring.
This ring forms after the companion star has passed through periastron; as dense wind leaves the donor star it compresses the accretion wake into a ring.
As the companion star moves in its orbit towards apastron, the ring moves away from the binary. 
In model Q2e08, because of the large ratio of wind velocity to instantaneous orbital velocity at apastron, little interaction between the wind and the companion star occurs at this distance, i.e. the outflow remains almost spherically symmetric and the accretion wake resembles the Bondi-Hoyle-Lyttleton case \citep{Bondi+Hoyle1944, Hoyle+Lyttleton1939}.   

Models Q2e04 and Q2e06 show an accretion disk around the companion star which builds up after the stars have passed periastron, but it disappears as the companion star approaches apastron. 
Two numerical effects may be contributing to this behaviour: on the one hand, similar to the circular model in Paper I, the radius of the accretion disk varies over time. 
Although we set the sink radius to be small, it can happen that the radius of the disk becomes smaller than the sink, so that the disk is engulfed by the sink and its mass is added to the accretor star. 
On the other hand, when the stars are at apastron little interaction between the gas and the stars takes place and the gas density remains low. 
With the low resolutions chosen, the typical smoothing length of the gas particles at apastron can become larger than the sink radius, which can cause  numerical artefacts near the accretor. 

Model Q2e06v1 (top row of Figure \ref{p3:fig:density_q2_v1e5}) shows that when the initial wind velocity is low ($v_\mathrm{init} = 1$ km s$^{-1}$) the geometry of the outflow becomes more complex, implying a stronger interaction between the wind and the companion star. 
In this model we observe a dense ring tightly wound around the binary, which builds up as the companion star moves through periastron. 
Although at the scale displayed in Figure \ref{p3:fig:density_q2_v1e5} it cannot be observed, this model shows a stream of gas flowing from the donor star towards the companion star which resembles wind Roche-lobe overflow \citep{val-borro, shazrene1, shazrene}, which is not observed in model Q2e06.
This gas stream is formed as the stars approach periastron and it vanishes as the companion star makes it way towards apastron. 
A similar mass transfer mechanism was found in Paper II for models in which the donor star was in corotation and $v_\mathrm{init} \lesssim 5$ km s$^{-1}$. 
Model Q2e06v1 also shows an accretion disk that forms at the passage through periastron, but as explained above it is engulfed by the sink after the passage through apastron. 

Model MMe05 shows the most complex morphology among our models (see the bottom panels of Figure \ref{p3:fig:density_q2_v1e5}). In this model the mass ratio is similar to models Q2e0 to Q2e08, but the stars are more massive, the donor star has a smaller radius, and it is non-rotating. 
An accretion disk is also formed, but contrary to the previously discussed models it is not engulfed by the sink although its size decreases as the companion star moves through periastron. 
In addition, contrary to the large eccentricity models in which the whole accretion wake is compressed in a ring, in this model two rings are observed.
A complete ring which surrounds the donor star is formed by the inner part of the accretion wake which builds up as the companion star moves through periastron. 
This inner ring is surrounded by an incomplete ring formed by the outer part of the accretion wake. 
The presence of both rings is probably related to the high density in the accretion wake in this model, which does not allow the wind to compress it into a single ring.
Another feature to notice in this model is that the external part of the wake is not clearly defined making the outflow very unsteady. 
In its circular counterpart (model V15a5, Paper I), we also observed a not-so-smooth accretion wake. 
However, compared to that model, the opening angle of the wake in model MMe05 is smaller and the accretion wake appears to be more misaligned with respect to the binary axis than in model V15a5. 
We should note, however, that a direct comparison between models MMe05 and V15a5 is hampered by the different assumptions in them.
One the one hand, in model V15a5 the velocity of the wind particles was forced to be constant, and the mass-loss rate was a factor of 15 smaller than in this work. 
On the other hand, the SPH resolution of the particles is much lower in this study, and the distance at which particles are removed from the simulation is smaller. 

Finally, we can attempt to compare models MMe05 and Q2e06v1 because their mass ratios are equal and their eccentricities are similar.
Furthermore, both models have a low wind-to-orbital-velocity ratio at periastron and a similar wind-to-orbital-velocity ratio at apastron (see Table \ref{p3:table:setup}). 
From Figure \ref{p3:fig:density_q2_v1e5} we observe that although both systems show a complex geometry, model MMe05 shows more irregularities in the outflow. 
When the stars are at periastron, the acretion wake of model MMe05 shows a much wider opening angle than model Q2e06v1. 
However, the opposite occurs when the stars are at apastron, i.e. at this point in the orbit the accretion wake of model Q2e06v1 is wider than in model MMe05. 
Additionally, at apastron the accretion wake of model MMe05 is much denser than for Q2e02v1 and very misaligned with respect to the binary axis. 
Model MMe05 creates the impression that the material the accretor star collects during its passage through periastron still interacts strongly with the accretor star when it reaches apastron. 
On the other hand, in model Q2e06v1 it appears from the density in the accretion wake that the strongest interaction between the gas and the stars occurs at periastron. 
We note that given the stellar parameters chosen for model MMe05, the temperature profile is somewhat different to that of model Q2e06v1. 
In order to check how this could affect the morphology of the outflow, we performed a test in which the effective temperature of the donor star in model MMe05 and the metallicity were similar to model Q2e06v1. 
However, no clear differences were found. 

\subsection{Mass-accretion rates}\label{p3:sec:only_beta}

\begin{figure*}
\centering
\includegraphics[width=0.95\hsize]{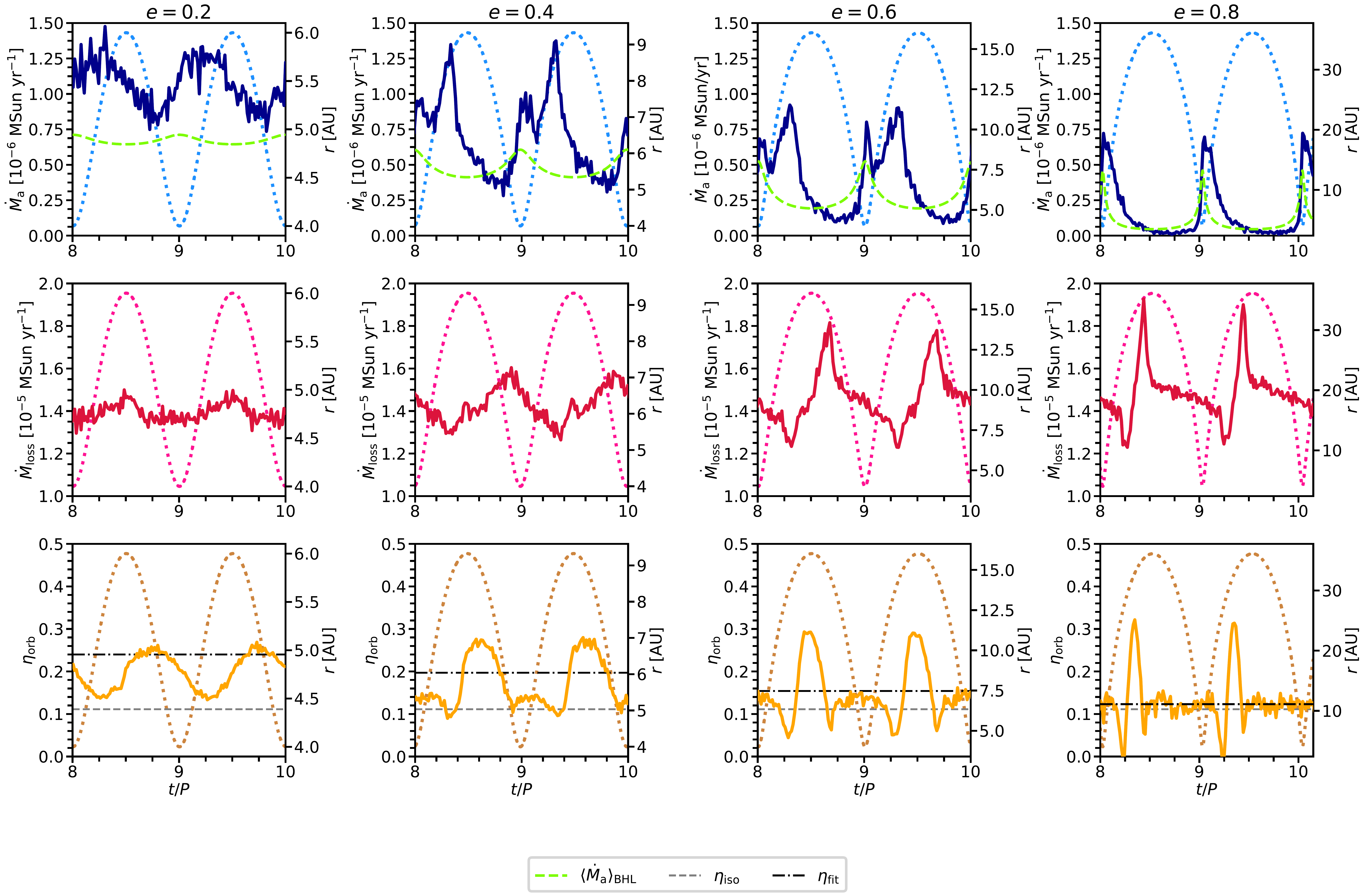}
\caption{\emph{Top:} Mass-accretion rate (solid dark blue line) as a function of the orbital phase, $t/P$, during the last two simulated orbits for models Q2a02 to Q2e08. The eccentricity increases from left to right. The green dashed line corresponds to the BHL accretion rate as computed with Eq. \ref{p3:eq:new_beta_BHL} and $\alpha_\mathrm{BHL} = 0.75$. The dotted light blue line corresponds to the distance between the stars (right-hand scale).
\emph{Middle:} Mass-loss rate as a function of time (solid red line) for the same models, measured as the flux crossing a sphere of radius $3a$. The dotted pink line corresponds to the distance between the stars.
\emph{Bottom:} Corresponding angular-momentum loss expressed by the parameter $\eta$ as a function of time (solid yellow line).
The distance between the star is shown with the dotted brown line. 
The dashed gray line corresponds to the isotropic-wind value $\eta_\mathrm{iso} = (1+q)^{-2}$, and the black dashed dotted line corresponds to the expected angular-momentum loss as computed by applying the fitting formula for angular-momentum loss from Paper II to the average orbital velocity.
}
\label{p3:fig:mdot_eta_q2s}
\end{figure*}

\begin{table}
\centering
\caption{Mass accretion efficiency}
\label{p3:table:beta}
\begin{tabular}{c c c}
\hline\hline
Model & $\langle\beta\rangle_\mathrm{BHL}$ & $\langle\beta\rangle_\mathrm{hydro}$\\ \hline \hline
Q2e0 & 0.056 & 0.094\\
Q2e02 & 0.045 & 0.071 \\
Q2e04 & 0.034  & 0.046 \\
Q2e06 & 0.023  & 0.026\\
Q2e08 & 0.011 & 0.010 \\
Q2e06v1 & 0.162 & 0.059 \\
MMe05 & 0.136  & 0.293\\
\hline
\end{tabular}
\tablefoot{
$\langle\beta\rangle_\mathrm{BHL}$ corresponds to the average mass-accretion efficiency for the BHL formalism, $\langle\dot{M}_\mathrm{a}\rangle/\dot{M}_\mathrm{d}$, with $\langle\dot{M}_\mathrm{a}\rangle$ from Eq. \ref{p3:eq:old_beta_BHL}. 
$\langle\beta\rangle_\mathrm{hydro}$ is the average mass-accretion efficiency per orbit obtained from the hydrodynamical models, computed over the last five orbits.
}
\end{table}

The Bondi-Hoyle-Lyttleton (BHL) formalism \citep{Hoyle+Lyttleton1939, Bondi+Hoyle1944, edgar} gives an estimate for the rate at which mass is accreted by a body moving in a gas medium. This model is often applied to wind accretion in binary systems, although the assumption of a uniform density and velocity field does not hold, especially for AGB winds. 
For binary stars in eccentric orbits the average mass-accretion rate is usually taken as \citep{boffin1988}:
\begin{equation}
\label{p3:eq:old_beta_BHL}
\langle\dot{M}_\mathrm{a}\rangle = -\alpha_\mathrm{BHL} \frac{\dot{M}_\mathrm{d}}{\sqrt{1-e^2}} \left(\frac{G M_\mathrm{a}}{a v_\mathrm{w}^2}\right)^2 \left[1+\left(\frac{v_\mathrm{orb}}{v_\mathrm{w}}\right)^2\right]^{-3/2},
\end{equation}
where $\alpha_\mathrm{BHL}\approx0.75$ is a constant\footnote{
Note that in \cite{boffin1988} $\alpha_\mathrm{BHL} = \alpha/2$, with $\alpha$ a constant between 1 and 2.},
$\dot{M}_\mathrm{d}<0$ is the rate at which the donor star is losing mass, $G$ is the gravitational constant, $a$ is the semi-major axis of the system, $v_\mathrm{w}$ is the local wind velocity, and $v_\mathrm{orb}^2 = G(M_\mathrm{d} + M_\mathrm{a})/a$ is the relative orbital velocity in a circular orbit with the same $a$. 
In an eccentric system the relative orbital velocity and the separation of the stars are time-variable parameters.
For this reason, in order to get a better estimate for the instantaneous mass-accretion rate, we substitute $a$ and $v_\mathrm{orb}$ in Eq. \ref{p3:eq:old_beta_BHL} by the instantaneous orbital separation, $r$, and instantaneous relative velocity of the stars, $v$, so that \citep{shazrene_thesis}:
\begin{equation}
\label{p3:eq:new_beta_BHL}
\dot{M}_\mathrm{a} = -\alpha_\mathrm{BHL} \frac{\dot{M}_\mathrm{d}}{r^2} \left(\frac{G M_\mathrm{a}}{v_\mathrm{w}^2}\right)^2 \left[1+\left(\frac{v}{v_\mathrm{w}}\right)^2\right]^{-3/2}.
\end{equation}

The top panels of Figures \ref{p3:fig:mdot_eta_q2s} and \ref{p3:fig:mdot_eta_q2e5_v1} show the mass-accretion rate onto the companion star as measured from the masses of the gas particles which cross the sink boundary per timestep. 
Given the discreteness of the SPH model, the mass-accretion rates show an associated shot noise. 
In order to suppress statistical fluctuations, we average the accreted mass over long time intervals (c.f. Paper I, section 3.5). 
For better appreciation, we only show the mass-accretion rate for two orbits of each model. The dotted lines in the figures show the distance between the stars for better recognition of the orbital phases. 
The green dashed lines overplotted in each figure show the BHL analytical estimate as computed from Eq. \ref{p3:eq:new_beta_BHL}.

For models Q2e02 to Q2e08 the BHL prescription predicts a peak in the mass-accretion rate when the stars are at their closest distance and a minimum in the mass-accretion rate at apastron. This occurs because in Eq. \ref{p3:eq:new_beta_BHL} the term containing the distance of the stars $r^{-2}$ dominates over the factor containing $v/v_\mathrm{w}$.
Likewise the BHL formalism predicts a decrease in the mass-accretion rate for large eccentricities. 
We observe a similar behaviour in our models.
Models Q2e04 and Q2e06 show an extra peak in the mass-accretion rate before the stars reach their maximum distance. 
As discussed in the previous section, an accretion disk builds up after the passage through periastron in these systems. 
Since the size of the disk is not constant, when the radius of the disk becomes smaller than the sink radius the material in the disk is swallowed by the sink, which is seen as an increase in the mass-accretion rate. 

Model Q2e06v1 (top panel of Fig. \ref{p3:fig:mdot_eta_q2e5_v1}) shows two peaks in the mass-accretion rate, one occurring at periastron and the second occurring near apastron. 
Similar to model Q2e06, in this model we also observe an accretion disk that builds up after the passage of the stars through periastron and is swallowed by the sink when the stars approach apastron. 
However, we note that for a system with the characteristics of model Q2e06v1 the BHL formalism predicts that the maximum in the accretion rate occurs when the stars are at their maximum separation, rather than at periastron, 
because due to the low wind velocity in this model the factor containing the term $v/v_\mathrm{w}$ dominates over the factor $r^{-2}$ in Eq.~\ref{p3:eq:new_beta_BHL}.  
Thus, even though the maximum in the mass-accretion efficiency in our simulation occurs because at the same time the accretion disk is engulfed by the disk, a maximum at apastron is likely to occur. 
For the same model we observe an enhancement in the mass-accretion rate when the stars are at periastron, which peaks nearly at the same value as the accretion rate at apastron. 
 
Model MMe05 (right panel of Fig. \ref{p3:fig:mdot_eta_q2e5_v1}) only shows a large peak at apastron. 
Similar to model Q2e06v1, the theoretical BHL model predicts that for a system with the characteristics of model MMe05 the maximum in the mass-accretion rate occurs at apastron. 
However, the peak mass-accretion rate found at apastron in our numerical models is a factor of $\approx 4$ larger than that predicted by the BHL approximation.
This is in contrast to model Q2e06v1, in which the largest peak in the accretion rate also occurs at apastron, but the peak value is a factor of $\approx 2$ lower than predicted by the BHL formalism.

Table \ref{p3:table:beta} shows the average mass-accretion efficiency per orbit, $\langle\beta\rangle = \langle\dot{M}_\mathrm{a}\rangle/\dot{M}_\mathrm{d}$, for our models. 
For comparison we also compute the average mass-accretion efficiency as estimated by Eq. \ref{p3:eq:old_beta_BHL}.
The mass-accretion efficiencies we find for models Q2e0 to Q2e08 are up to a factor of $\approx 1.7$ higher than predicted by the BHL formalism, 
with the largest difference occurring for the model with a circular orbit. 
As the eccentricity increases to 0.8, the average accretion rate approaches the BHL approximation.
This is consistent with our findings in Paper~II that for circular orbits Eq.~\ref{p3:eq:old_beta_BHL} quite accurately describes the accretion efficiency when the ratio $v_\mathrm{w}/v_\mathrm{orb} \gg 1$, but underestimates $\langle\beta\rangle_\mathrm{hydro}$ for lower velocity ratios because of the stronger interaction between the companion and the gas in these cases. For models Q2e0 and Q2e02, $v_\mathrm{w}$ is smaller than the relative orbital velocity $v$ at any point in the orbit (see Table~\ref{p3:table:setup}), but for the more eccentric models $v_\mathrm{w}/v > 1$ near apastron, where the binary spends most of its time. As already noted in Sect.~\ref{p3:sec:morphology}, the outflow morphology and accretion wake near apastron indeed resemble the BHL case in these models.
For model Q2e06v1 the average accretion efficiency is a factor of $\approx 2.7$ lower than in the BHL formalism, whereas for model MMe05 $\langle\beta\rangle_\mathrm{hydro}$ is more than a factor of two higher than $\langle\beta\rangle_\mathrm{BHL}$ (and a factor of $\approx 1.3$ larger than for its circular counterpart, model V15a5 in Paper I). This is surprising given the similar values of $v_\mathrm{w}/v$ along the orbit in these models, but is likely related to their very different outflow morphologies as a function of orbital phase (see Sect.~\ref{p3:sec:morphology} and Fig.~\ref{p3:fig:density_q2_v1e5}). We note that a similar result as for model Q2e06v1 was obtained in Paper II, where for very low initial wind velocities we found mass-accretion efficiencies below the BHL prescription.

\subsection{Angular-momentum loss}\label{p3:sec:am}

\begin{figure}
\centering
\includegraphics[width=0.95\hsize]{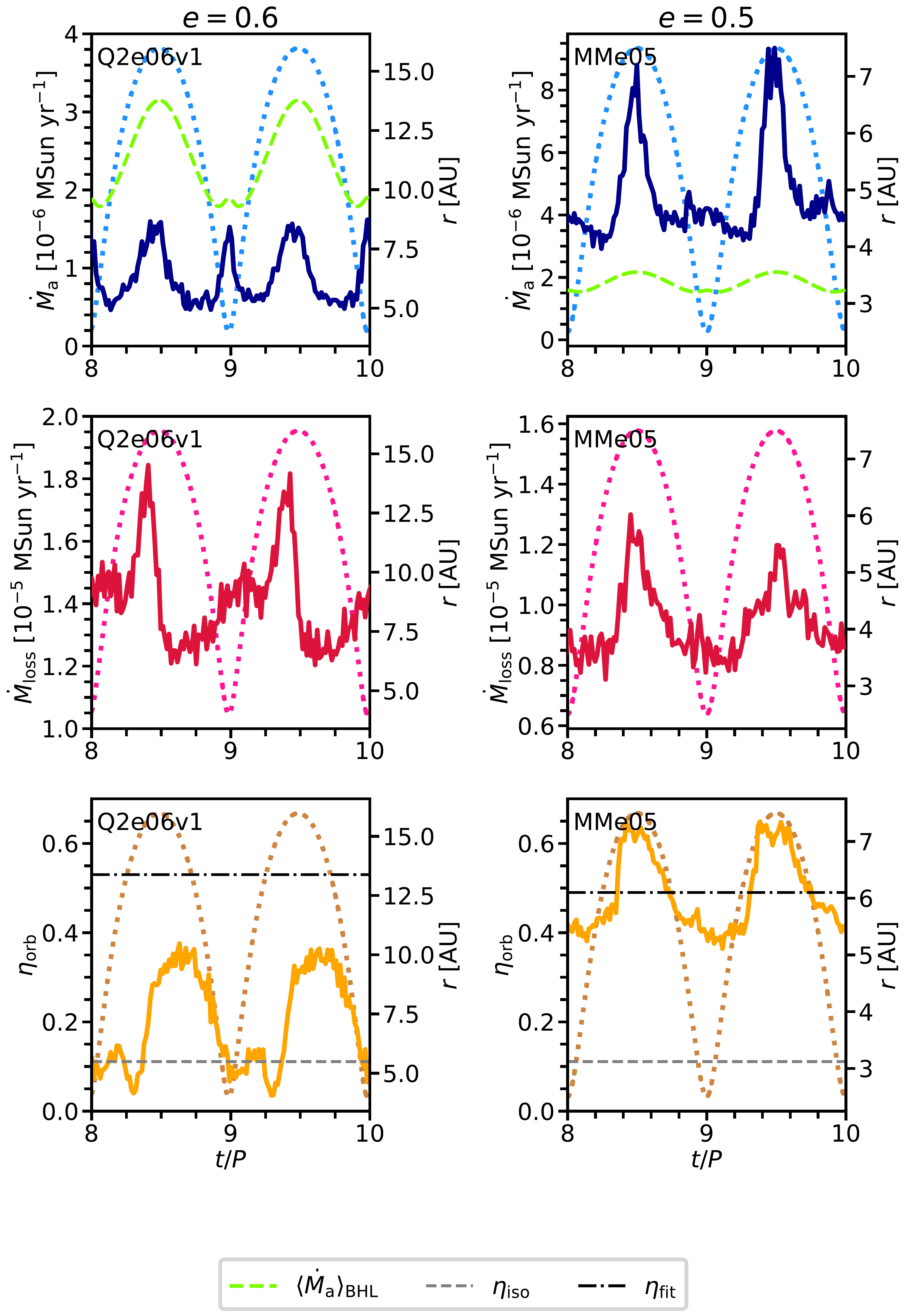}
\caption{Similar to Fig. \ref{p3:fig:mdot_eta_q2s}, but for models Q2e06v1 (left) and MMe05 (right). 
}
\label{p3:fig:mdot_eta_q2e5_v1}
\end{figure}

The middle panels of Figs. \ref{p3:fig:mdot_eta_q2s} and  \ref{p3:fig:mdot_eta_q2e5_v1} show the rate at which mass is lost from the binary system per orbit.  
We measure this quantity as the flux of mass crossing a sphere of radius $3a$.
We choose this radius because for circular orbits we have shown that beyond this distance no further exchange of angular momentum between the wind and the orbit takes place (see Paper I).  

The maximum in the mass-loss rate occurs at the time when most of the material in the ring (for highly eccentric systems) or in the spiral wake (for low-eccentricity systems) crosses the $3a$ boundary.
As gas moves away from the binary it removes angular momentum from the orbit, which was exchanged during the strong interaction at close orbital separations.
At the same time, gas also removes angular momentum due to the rotation of the donor star. 
Similar to Paper II, we parametrise the angular momentum lost as
\begin{equation}
\label{p3:eq:eta}
\dot{J} = \eta_\mathrm{orb} \frac{J_\mathrm{orb}}{\mu} (1 - \beta)\dot{M}_\mathrm{d} + \dot{J}_\mathrm{spin},
\end{equation}
where the first term on the right-hand side corresponds to the change in the orbital angular momentum, where $J_\mathrm{orb}$ is the orbital angular momentum of the binary, $\mu = M_\mathrm{d}M_\mathrm{a}/(M_\mathrm{d}+M_\mathrm{a})$ is the reduced mass of the binary, and $\beta$ is the average mass-accretion efficiency per orbit as computed from Eq. \ref{p3:eq:old_beta_BHL}.
The parameter $\eta_\mathrm{orb}$ is a dimensionless measure of the specific angular momentum taken from the orbit and transferred to the outflowing gas.
The second term in Eq. \ref{p3:eq:eta} is the contribution from the loss of spin angular momentum, which we take as $\dot{J}_\mathrm{spin} = \frac{2}{3} R_\mathrm{d}^2\dot{M}_\mathrm{d}\Omega_{\mathrm{orb}, pe}$. As shown in Paper II, this accurately describes the angular-momentum loss of a single AGB star in our simulations.
The angular momentum in the outflow, $\dot{J}$ on the left-hand side of Eq.~\ref{p3:eq:eta}, is measured at the time the SPH particles cross the $3a$ boundary.  
We only take the perpendicular component to the orbital plane of the angular momentum, $J_z$, since we have verified that the other two components, $J_x$ and $J_y$, are very small, i.e. the flow is symmetric with respect to the orbital plane. 

The bottom panels of Figs. \ref{p3:fig:mdot_eta_q2s} and  \ref{p3:fig:mdot_eta_q2e5_v1} show the specific angular momentum that the wind takes away from the orbit in terms of the parameter $\eta_\mathrm{orb}$. 
Note that due to the eccentricity of the systems the angular-momentum lost over one orbit is not constant.  
We compare the angular-momentum loss of our eccentric models to the isotropic-wind mode case, $\eta_\mathrm{iso} = (1+q)^{-2}$, and to the fitting formula for the specific-angular momentum loss, $\eta_\mathrm{fit} = \eta_\mathrm{orb}(q, v_\mathrm{w}/v_\mathrm{orb})$, obtained in Paper II. 
Since $v_\mathrm{orb}$ varies over the orbit, in order to apply the formula for $\eta_\mathrm{fit}$ we take $v_\mathrm{orb} = \langle v\rangle_\mathrm{orb}$, where $\langle v\rangle_\mathrm{orb}$ is the time-averaged velocity over the orbit which is dominated by the long time the stars spend near apastron.
Table \ref{p3:table:eta} shows the values for $\eta_\mathrm{iso}$, $\eta_\mathrm{fit}$, and the average specific angular momentum lost per orbit, $\langle\eta\rangle_\mathrm{orb}$, which in accordance with Eq. \ref{p3:eq:eta} is computed as 
\begin{equation}
\label{p3:eq:eta_avg}
\langle \eta\rangle_\mathrm{orb} =  \left(\frac{J_\mathrm{orb}}{\mu}\right)^{-1}\left[ \left(\frac{\sum_i^N J_{z,i}}{N m_g}\right)_\mathrm{orb} - \frac{2}{3} \frac{R_\mathrm{d}^2\Omega_{\mathrm{orb}, pe}}{(1-\beta)}\right],
\end{equation}
where $J_{z,i}$ is the perpendicular component of the angular momentum of the $i$'th particle which crosses the $3a$ boundary, $N$ is the number of particles that cross the $3a$ boundary in one orbit, and $m_\mathrm{g}$ is the mass of the SPH particles.

\begin{table}
\centering
\caption{Angular-momentum loss}
\label{p3:table:eta}
\begin{tabular}{c c c c}
\hline\hline
Model & $\eta_\mathrm{iso}$ & $\langle\eta\rangle_\mathrm{fit}$ & $\langle\eta\rangle_\mathrm{orb}$\\ \hline \hline
Q2e0 & 0.111 & 0.278   & 0.217 \\
Q2e02 & 0.111 & 0.240 & 0.199 \\
Q2e04 & 0.111 & 0.197 & 0.175 \\
Q2e06 & 0.111 & 0.154 & 0.151 \\
Q2e08 & 0.111 & 0.123 & 0.130 \\
Q2e06v1 & 0.111 & 0.411 & 0.198 \\
MMe05 & 0.111 & 0.402 & 0.494 \\
\hline
\end{tabular}
\tablefoot{
$\eta_\mathrm{iso}$ corresponds to the specific angular-momentum loss in units of $J/\mu$ for the isotropic-wind case. 
$\langle\eta\rangle_\mathrm{fit}$ corresponds to the average angular-momentum loss as derived by computing the average orbital velocity during one orbit and applying the fit for $\eta(q, v_\mathrm{w}/v_\mathrm{orb})$ obtained in Paper II. 
$\langle\eta\rangle_\mathrm{orb}$ is the average angular-momentum loss per orbit for the numerical models presented in this paper, computed over the last five orbits. 
}
\end{table}

\begin{figure*}
\centering
\includegraphics[width=0.9\hsize]{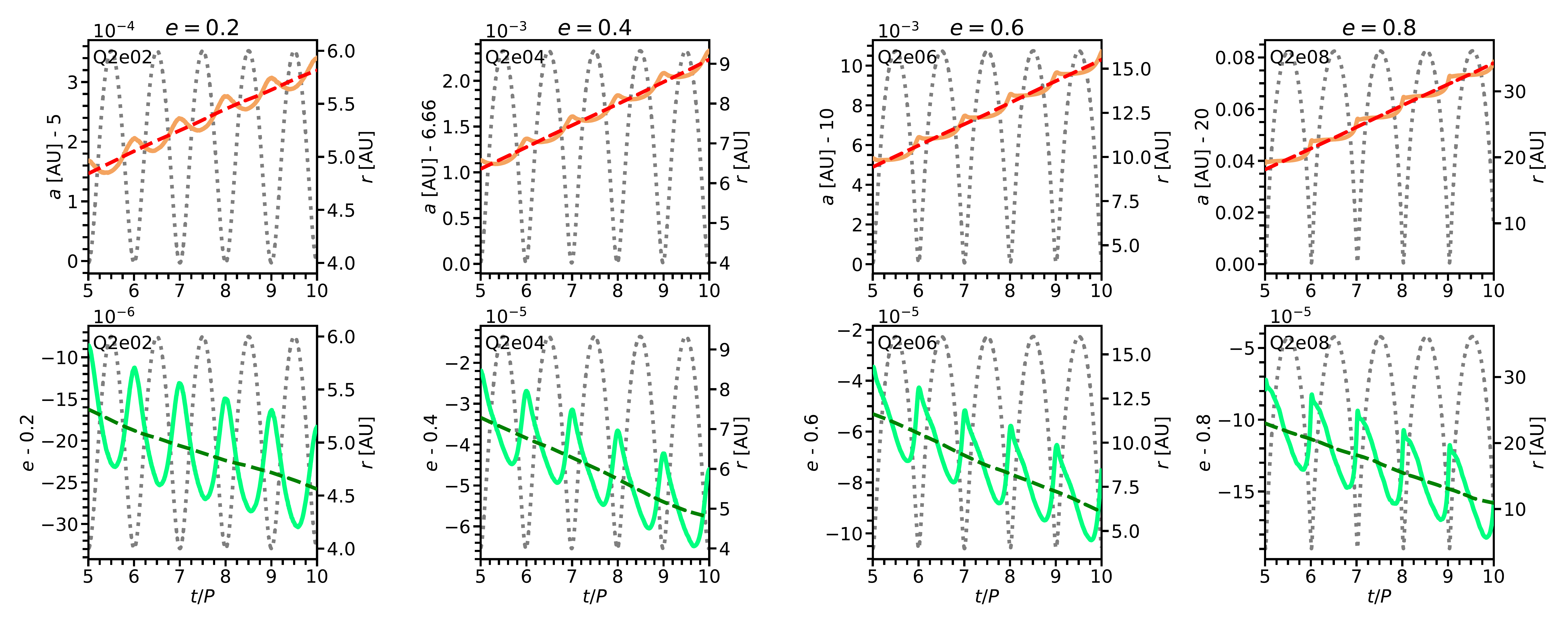}
\caption{\emph{Top:} Semi-major axis as a function of time, relative to the orbital period, for models Q2e02 to Q2e08 (solid lines). 
Notice that since the change in the orbit is very small (between $10^{-4}$ and $10^{-2}$ AU), for better visualisation the order of magnitude of the quantities along the $y$-axis is given on the top left corner.
\emph{Bottom:} The eccentricity as a function of $t/P$ for models Q2e02 to Q2e08 (solid lines). 
Similar to $a$, the order of magnitude of the quantities along the $y$ axis is given on the top left corner.
The dashed lines in each panel show the moving average of $a$ and $e$ over the interval $[t/P-0.5, t/P+0.5]$, for better appreciation of the long-term trend. We only show the evolution of $a$ and $e$ for the last five orbits of our simulations. 
The dotted gray lines correspond to the distance between the stars, with the minimum corresponding to periastron and the maximum to apastron.}
\label{p3:fig:a_e_q2models}
\end{figure*}

For models with similar stellar parameters but different eccentricities (Q2e02 to Q2e08), we observe that as the eccentricity increases the angular-momentum loss decreases. 
For the mildly eccentric model Q2e02 the loss in angular momentum is almost the same as in the circular case. 
However, for the most eccentric case Q2e08 the angular-momentum loss is smaller by a factor of $\approx 1.5$ and approaches the isotropic-wind mode. 
This is not surprising since the companion star spends most of its time at apastron, where the outflow is not strongly modified and has an almost spherically symmetric geometry, as shown by the plot of the gas density in the orbital plane for this model (bottom panels of Fig. \ref{p3:fig:density1}).
Model Q2e06v1 shows an $\eta$ value a factor of $\approx 1.3$ larger than its counterpart with larger wind velocity, which reflects the stronger interaction between the wind and the companion star occurring in this model. 

If we compare models MMe05 and Q2e06, which have nearly the same eccentricity and equal mass ratios, we notice that in model MMe05 a much larger amount of angular momentum is lost. 
We recall that in model MMe05, the geometry of the outflow is strongly modified compared to model Q2e06 (see Sect. \ref{p3:sec:morphology}). 
In model MMe05 a very dense accretion wake is observed behind the companion star during the whole orbit at an angle that is considerably misaligned with the binary axis. 
An accretion wake with these characteristics (high density and misalignment with the binary axis) will exert a stronger torque on the companion star allowing a larger exchange of angular momentum between the wind and the orbit of the stars.
Finally and although it cannot be observed in Figure \ref{p3:fig:mdot_eta_q2e5_v1}, the angular-momentum loss in model MMe05 is increasing as a function of time. 
It is not clear why this increase occurs, but it suggests that at each passage of the stars through periastron a stronger interaction takes place. 

Finally, we find that $\langle\eta\rangle_\mathrm{orb}$ agrees within $\approx 20\%$  to the fit given in Paper II for $\eta(q, v_\mathrm{w}/\langle v\rangle_\mathrm{orb})$ for models Q2e0-Q2e08 and MMe05. 
The best agreement occurs for large eccentricities ($e=0.6$ and $e=0.8$).
However, for model Q2e06v1 we find that the orbital angular momentum lost from the system is a factor $\approx 2$ lower than what our fitting formula predicts.

\subsection{Changes in the orbital elements}\label{p3:sec:orb}

Because we use an N-body code to compute the dynamics of the stars, it is possible to measure the change in the semi-major axis and eccentricity of the orbit directly from the simulations within the numerical error (see Paper II). 
The total energy per reduced mass, $\epsilon$, of two bodies orbiting each other under the influence of their gravity is given by the sum of the kinetic energy and the potential energy of the system,
\begin{equation}
\label{p3:eq:energy_bodies}
\epsilon = \frac{v^2}{2} - \frac{G(M_\mathrm{d}+M_\mathrm{a})}{r}.
\end{equation}
Here $r$ and $v$ are the magnitudes of the relative position and velocity vectors of the stars, $\mathbf{r}$ and $\mathbf{v}$, which we measure from the simulation. In addition, the angular momentum per reduced mass of the system can be written as:
\begin{equation}
\label{p3:eq:am_bodies}
\ell = |\mathbf{r} \times \mathbf{v}|.
\end{equation}
On the other hand, the orbital energy per reduced mass and specific angular momentum of a system in a Keplerian orbit are given by 
\begin{equation}
\label{p3:eq:energy_ellipse}
\epsilon = -\frac{G(M_\mathrm{d}+M_\mathrm{a})}{2a} 
\end{equation}
and 
\begin{equation}
\label{p3:eq:am_ellipse}
\ell = G (M_\mathrm{d}+M_\mathrm{a})\sqrt{\frac{e^2-1}{2\epsilon}}.
\end{equation}
By combining Eqs. \ref{p3:eq:energy_bodies} and \ref{p3:eq:energy_ellipse}, we can determine $a$ at any given time in our simulations. 
A similar calculation can be done for $e$ by combining Eqs. \ref{p3:eq:am_ellipse}, \ref{p3:eq:energy_ellipse} and \ref{p3:eq:am_bodies}.

\begin{figure}
\centering
\includegraphics[width=0.9\hsize]{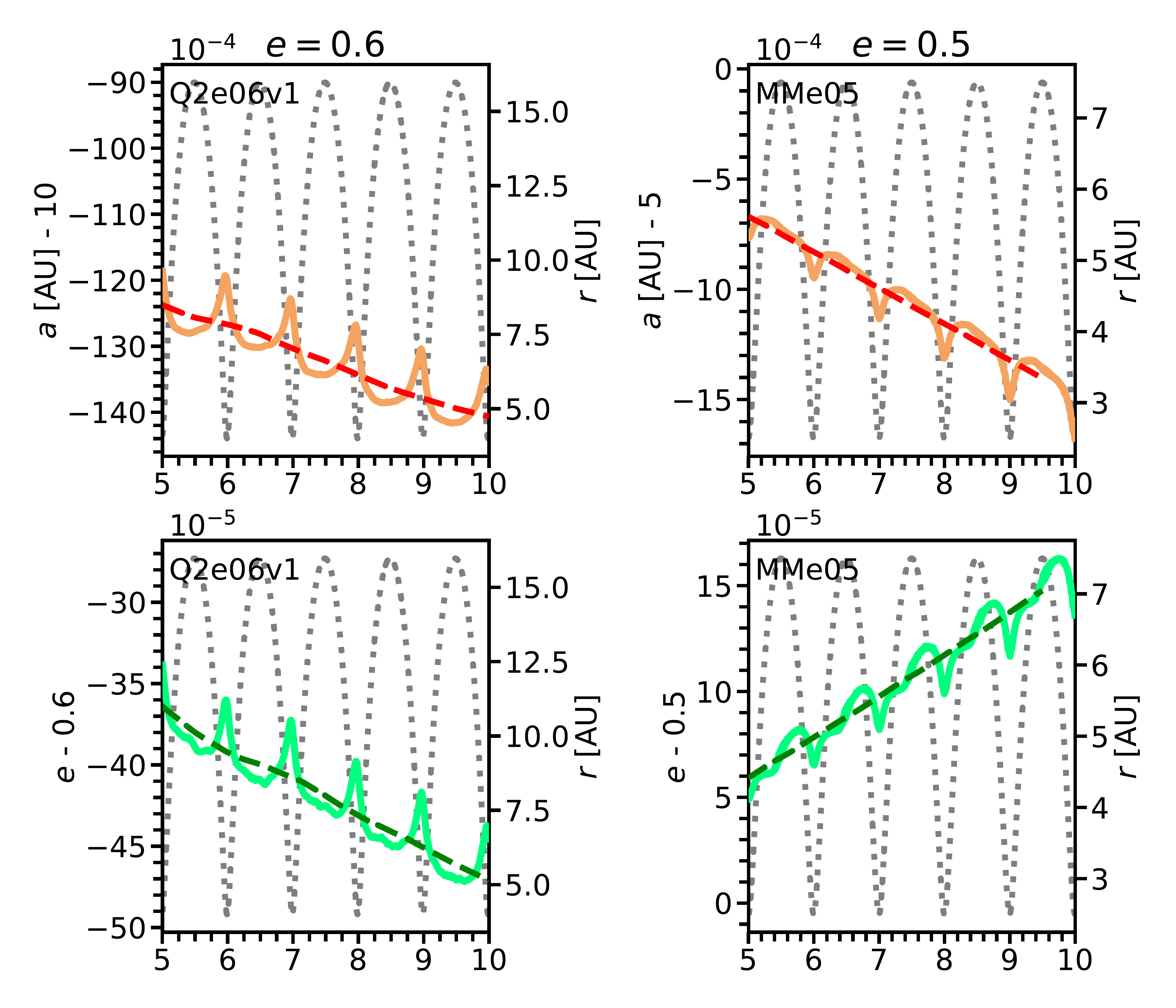}
\caption{Similar to Fig. \ref{p3:fig:a_e_q2models}, but for models Q2e06v1 (left) and MMe05 (right).}
\label{p3:fig:a_e_q2_diff}
\end{figure}

Figs. \ref{p3:fig:a_e_q2models} and \ref{p3:fig:a_e_q2_diff} show the evolution of $a$ and $e$, as computed above, for the last five orbits of our simulations. 
We assume that similar to Paper I, where we simulated circular orbits, after the fourth orbit a quasi-steady state is reached. 
For models Q2e02 to Q2e08, in which angular-momentum loss is relatively small, a similar long-term trend for $a$ and $e$ is observed where the semi-major axis increases during the evolution of the system and the eccentricity decreases. 
We note that the short-timescale variations seen in Figs.~\ref{p3:fig:a_e_q2models} and \ref{p3:fig:a_e_q2_diff} have little physical meaning, because $a$ and $e$ are only well-defined for a complete orbit.
Models Q2e06v1 and MMe05 show the opposite trend in $a$ to models Q2e02 to Q2e08.  
Since more angular momentum is lost in these systems their orbits are seen to be shrinking. 
However, model Q2e06v1 shows a decrease in eccentricity, whereas in model MMe05 the eccentricty increases. 
This likely results from the location along the orbit where most of the angular momentum exchange takes place. 
In model Q2e06v1 (and the other Q2e0$i$ models) it appears from the high density in the accretion wake at periastron that the exchange of angular momentum mainly occurs while the stars are near periastron, which will result in a decrease in eccentricity since the companion star will be slowed down at periastron. 
However, if the strongest torque occurs at apastron (as appears to be the case for model MMe05 from Figure \ref{p3:fig:density_q2_v1e5}), the eccentricity will be pumped since the companion star will be slowed down before it moves through periastron again. 

The change in orbital separation and eccentricity over a time interval $\Delta t$ can be derived from Eqs. \ref{p3:eq:energy_ellipse} and \ref{p3:eq:am_ellipse}, yielding:
\begin{equation}
\label{p3:eq:da_a}
\frac{\Delta a}{a} = -\frac{\Delta \epsilon}{\epsilon} + \frac{\Delta M}{M},
\end{equation} 
where $M = M_\mathrm{d} + M_\mathrm{a}$, and
\begin{equation}
\label{p3:eq:de_e}
\frac{\Delta e}{e} = \frac{1-e^2}{e^2} \left(\frac{1}{2} \frac{\Delta a}{a} + \frac{1}{2} \frac{\Delta M}{M} - \frac{\Delta \ell}{\ell}\right).
\end{equation}
The change in $\epsilon$ and $\ell$ is derived from the relative velocity and separation of the stars, which we know at any time in our simulations (Eqs. \ref{p3:eq:energy_bodies}  and \ref{p3:eq:am_bodies}). 
We estimate $\dot{a}$ and $\dot{e}$ as the average rate of change given by Eqs. \ref{p3:eq:da_a} and \ref{p3:eq:de_e} over the last five orbits. 
The resulting values of $\dot{a}/a$ and $\dot{e}/e$ are shown in Table \ref{p3:table:da_a_de_e}. 
For models Q2e02 to Q2e08, we observe that as the change in angular momentum approaches the isotropic regime (for systems with high eccentricities) the systems widens at a faster rate. 
For instance, over the ten orbits we run our simulations, in model Q2e08 the orbit expands by about $0.08$ AU.
On the other hand, systems with small eccentricities (Q2e02 and Q2e04) widen slowly. 
In their circular counterpart (model Q2e0), we find that the system shrinks due to the relatively large amount of angular momentum lost. 

In model MMe05 the orbit shrinks very fast compared to the rest of the models, and the eccentricity increases at a high rate. This is partly a consequence of our choice of the mass-loss rate, which is similar to the other models even though the stellar parameters for this system are different.
As the donor star in this model has a smaller radius it should actually lose mass at a slower rate. 
We have verified that by setting a lower mass-loss rate ($10^{-6}$ M$_\odot$ yr$^{-1}$, as in Paper I) the angular-momentum loss rate and the mass-accretion rate scale down proportionally,
i.e. the rate at which the semi-major axis and eccentricity change per \emph{unit mass lost} is similar. 

\subsubsection{Comparison to analytical results}

\begin{table*}
\centering
\caption{Change in the orbital elements}
\label{p3:table:da_a_de_e}
\begin{tabular}{c c c c c c}
\hline\hline
Model & $\dot{a}/a$ & $(\dot{a}/a)_\mathrm{DK16}$ & $\dot{e}/e$ & $(\dot{e}/e)_\mathrm{DK16}$ & $(\dot{e}/e)_\mathrm{tides}$\\ 
- & yr$^{-1}$ & yr$^{-1}$ &  yr$^{-1}$ &  yr$^{-1}$ & yr$^{-1}$ \\ \hline \hline
Q2e0 & $-8.43\times 10^{-7}$ & $5.20\times 10^{-6}$ &  - & - & -  \\
Q2e02 & $8.28 \times 10^{-7}$  & $5.77\times 10^{-6}$  & $-1.18\times 10^{-6}$ & $-2.40\times 10^{-6}$ &  $-1.10 \times 10^{-4}$  \\
Q2e04 & $2.77 \times 10^{-6}$ & $6.22\times 10^{-6}$  & $-9.58 \times 10^{-7}$ & $-1.53\times 10^{-6}$ & $-1.91 \times 10^{-5}$\\
Q2e06 & $4.56 \times 10^{-6}$  & $6.49\times 10^{-6}$ & $-5.37 \times 10^{-7}$ & $-8.67 \times 10^{-7}$ & $-4.41 \times 10^{-6}$\\
Q2e08 & $6.15 \times 10^{-6}$ & $6.81\times 10^{-6}$ & $-2.16 \times 10^{-7}$ & $-3.33 \times 10^{-7}$ & $-9.22 \times 10^{-7}$\\
Q2e06v1 & $-1.58\times 10^{-6}$& $4.15 \times 10^{-6}$ & $-1.66 \times 10^{-6}$ & $-1.97 \times 10^{-6}$ & $-4.41 \times 10^{-6}$\\
MMe05 & $-6.88 \times 10^{-6}$ & $-3.18\times 10^{-6}$ & $6.41 \times 10^{-6}$ & $-3.91 \times 10^{-6}$ &$-5.10 \times 10^{-6}$ \\
\hline
\end{tabular}
\tablefoot{
$\dot{a}/a$ and $\dot{e}/e$ correspond to the changes in semi-major axis and eccentricity, respectively, as measured dynamically from the simulations. 
The values of $\dot{a}/a$ and $\dot{e}/e$ with the subscript DK16 correspond to the analytical values for these quantities as derived from Eqs. 81 and 82 in \cite{Dosopoulou+Kalogera2016}.
$(\dot{e}/e)_\mathrm{tides}$ is the expected change in the eccentricity due to tidal interaction, as derived from Eq. \ref{p3:eq:Hut_eq10}.
}
\end{table*}

\cite{Eggleton2006} and \cite{Dosopoulou+Kalogera2016} have derived analytical expressions for the secular evolution of $a$ and $e$ under the assumption of mass loss by fast isotropic winds. In Table \ref{p3:table:da_a_de_e} we compare our simulation results to the expressions for $\dot{a}/e$ and $\dot{e}/e$ of \citet[eqs. 81 and 82]{Dosopoulou+Kalogera2016}. 
To make the comparison we replace the assumed Bondi-Hoyle accretion rate in these equations with the average mass-accretion rate obtained from our hydrodynamical models.
Our results for models Q2e02 to Q2e08 show an agreement in the sign of $\dot{a}$ and $\dot{e}$, but the magnitude of both quantities is smaller than predicted by the analytical model. The orbit expands less rapidly, by a factor between about 1.1 and 7, and the eccentricity decreases at a lower rate, by a factor between about 1.4 and 2.0.  In both cases, the largest difference occurs for the least eccentric model, Q2e02, and as noted earlier in the circular case Q2e0 the orbits even shrinks rather than expands. This trend is understandable from the fact than in our most eccentric models, during the relatively long time spent near apastron the outflow is nearly isotropic and relatively fast compared to the orbital speed (see Sect.~\ref{p3:sec:morphology}), corresponding to what is assumed in the analytical model of \cite{Dosopoulou+Kalogera2016}. On the other hand, for smaller eccentricities the outflow is more strongly modified, leading to larger differences with the analytical expressions.

For model Q2e06v1 the analytical model predicts slower expansion and faster circularisation than for model Q2e06, as a result of the higher average mass-accretion rate (Table~\ref{p3:table:beta}). Our hydrodynamical results show that the orbit shrinks rather than expands, and as for the other Q2e models the eccentricity decreases somewhat more slowly than predicted. The largest difference with the analytical model occurs for MMe05, where we find $\dot{e} > 0$ as previously noted. The equations of \cite{Dosopoulou+Kalogera2016} instead predict a fairly rapid decrease of the eccentricity, and slower shrinkage of the orbit by approximately a factor of two. We note that in the analytical model an increase in eccentricity only occurs for binaries with $0 < q < 0.78$.
These differences can again be ascribed to the strong modification of the outflow in these two models, compared to the fast isotropic wind case. The comparison in this section indicates that, in general, stronger interaction between the outflow and the binary leads to (1) less orbital expansion or faster orbital shrinking, and (2) slower circularisation or, in extreme cases, an increase in eccentricity.

\section{Discussion}\label{p3:sec:discussion}

In this section, we discuss how the geometry of the outflow in our hydrodynamical models compares with observations of AGB binary systems which are thought to be in eccentric orbits, and how our exploratory results for the rate of change of the eccentricity compare to the tidal circularisation timescales. 
In addition, we briefly discuss some of the numerical and physical effects that may affect our results, and some future work that could be performed in order to gain a better understanding of how wind mass transfer can impact the eccentricity of binary systems with a red-giant component.

\subsection{Comparison to observations}

Our work shows that wind mass transfer in eccentric binaries results in geometries of the outflow which differ considerably from the circular binary case when $e \gtrsim 0.5$.
The present study thus confirms the potential of numerical models for constraining the eccentricity in observed interacting binary stars.
For instance, an outflow morphology which shows a disrupted ring has been observed in the inner region of the spiral pattern of the binary system AFGL 3068.
Hydrodynamical models have shown that the geometry of this system can be reproduced if the eccentricity of the system is about 0.8 \citep{Kim+2017}.
However, the bifurcation in the spiral described by these models and also observed in AFGL 3068 is not found in our work. 
We note that a direct comparison between our models and the observations of AFGL 3068 cannot be made,
because the spiral pattern of this object extends up to $\approx 60$ times the mean orbital separation of the system, whereas we remove particles at much shorter distances from the binary.
Furthermore, the mass ratio of AFGL 3068 is different the value assumed in our work. 
Another system for which incomplete ring patterns have been observed is the carbon star CIT 6, which is believed to contain a star evolving from the AGB to the post-AGB phase \citep{Kim+2013}.
To explain the observed geometry of the outflow of CIT 6, a binary companion with a very high eccentricity has been suggested \citep{Kim+2015}.
However, for similar reasons to those mentioned above (different mass ratios and the fact that we remove particles close to the binary) no direct comparison can be made between our models and these observations. 
 
\subsection{Orbital evolution timescales}

Our results shed some light on the evolution of the orbital parameters when wind mass transfer occurs. 
However, we note that the orbital evolution is also affected by physical processes that are not included in our models.
For instance, given the large sizes of AGB stars, for close binary systems circularisation of the obit is likely to occur due to tidal effects.
Since the stars in our models are approximated by point particles, this effect is not taken into account. 

In order to estimate the circularisation timescale predicted by tidal evolution, we use eq. 10 from \cite{Hut1981}: 
\begin{equation}
\label{p3:eq:Hut_eq10}
\begin{aligned}
\frac{\dot{e}}{e} = & -27 \left(\frac{k}{T}\right)q^{-1}(1+q^{-1})\left(\frac{R_\mathrm{d}}{a}\right)^8\frac{1}{(1-e^2)^{13/2}}  \\
 & \times \left[f_3(e^2) - \frac{11}{18} (1-e^2)^{3/2} f_4(e^2)\frac{\Omega_\mathrm{spin}}{\Omega_\mathrm{bin}}\right], 
\end{aligned}
\end{equation}
where $k$ is the apsidal motion constant of the donor star, $T$ is the time-scale on which significant changes in the orbit take place through tidal evolution, $f_3$ and $f_4$ are polynomial functions of $e^2$ given by \cite{Hut1981}, $\Omega_\mathrm{spin}$ is the angular velocity of the donor star and $\Omega_\mathrm{bin} = 2\pi/P$ is the mean angular velocity of the binary. 
We take $(k/T)$ as in eq. 30 from \cite{hurley}, with the mass of the envelope equal to $M_\mathrm{env} = 0.55$ M$_\odot$ in models where $M_\mathrm{d} = 1.2$ M$_\odot$, and $M_\mathrm{env} = 2.4$ M$_\odot$ for $M_\mathrm{d} = 3$ M$_\odot$. 
For an AGB star the size of the core is negligible compared to the convective envelope, thus we approximate the radius of the envelope as $R_\mathrm{env} = R_\mathrm{d}$.
Note that in Eq. \ref{p3:eq:Hut_eq10} the sign of $\dot{e}$ is determined by the last factor containing $\Omega_\mathrm{spin}/\Omega_\mathrm{bin}$, i.e. $\dot{e} > 0$ is possible but only for sufficiently fast rotation \cite[see][for a discussion]{Hut1981}. 

Table \ref{p3:table:da_a_de_e} shows our estimates for the circularisation timescales for systems with binary parameters as in our simulations. 
For models Q2e02 and Q2e04, tidal circularisation is much more effective than the circularisation induced by wind interaction, whereas for models with $e\gtrsim0.6$ the tidal circularisation timescale is similar to the hydrodynamical circularisation timescale.
We can roughly estimate by how much the eccentricity of these models will decrease by the time the star leaves the AGB phase. 
A star with the characteristics of our donor star will spend another $\approx 3 \times 10^{4}$ yr in the superwind phase before it leaves the AGB (see Paper II for the method used to evolve this star). 
By assuming that the tidal circularisation timescale is constant during this time interval, we find that by the time the star leaves the AGB phase the binary will have an eccentricity of $\approx 0.007$ for model Q2e02 and $e \approx 0.23$ for model Q2e04.
However, for models Q2e06 and Q2e08, as well as Q2e06v1, the circularisation timescales of both tidal interaction and wind interaction are so long that the change in the eccentricity before the donor star leaves the AGB will be very small.

For model MMe05, our hydrodynamical models predict an increase in eccentricity. 
In this case, the tidal circularisation timescale is of the same order of magnitude as the hydrodynamical eccentricity pumping timescale.
Therefore it is possible that these effects counteract each other, leading only to a small change in the eccentricity.
However, as mentioned in Sect. \ref{p3:sec:orb} the assumed mass-loss rate of this system is higher than expected for the superwind phase of an AGB star with the stellar parameters of model MMe05. 
By assuming a lower mass-loss rate for this model, $\dot{M}_\mathrm{d}=10^{-6}$ M$_\odot$ yr$^{-1}$ (as in Paper I), we find that $\dot{e}/e = 4.55 \times 10^{-7}$ yr $^{-1}$. 
In that case, the eccentricity pumping timescale from the hydrodynamical models will not be able to compete with the tidal circularisation timescale. 
Furthermore, since the semi-major axis is decreasing in this system, tidal forces will become stronger as the system evolves. 
From Eqs.~\ref{p3:eq:Hut_eq10} and \ref{p3:eq:omegaper}, we see that tidal evolution would only be able to pump the eccentricity in a system like MMe05 if $\Omega_\mathrm{spin} > 1.1 \Omega_{\mathrm{orb}, pe}$, where $\Omega_{\mathrm{orb}, pe}$ is the angular velocity at periastron. 

\subsection{Model MMe05}

Model MMe05 shows that it is possible to find a regime where the eccentricity increases due to wind mass transfer, and this trend is quite robust against various tests we have performed.
In a test simulation with a sink radius that is twice as large (i.e., equal to $0.2 R_{\mathrm{L,2}|pe}$), we find that the average mass-accretion efficiency increases compared to the value given in Table \ref{p3:table:beta}, but the angular-momentum loss remains constant. This also results in an increase of the eccentricity and a decrease of the semi-major axis.
We have also verified that regardless of the assumed mass-loss rate of the donor star 
and the temperature profile, the results always lead to an increase in eccentricity.
However, there are many other characteristics of this model which make it difficult to compare to the more realistic models Q2e0$i$.
On the one hand, unlike models Q2e0$i$ where the stellar parameters of the donor were taken from a stellar evolution code, the stellar parameters of the AGB star for model MMe05 were chosen arbitrarily to match the parameters of the systems studied in Paper I. 
Furthermore, in order to compare our results for this model to its circular counterpart from Paper I, we neglected the possibility of pseudo-synchronisation of the donor star. 
As seen in Paper II, rotation of the donor star for low $v_\mathrm{w}/v_\mathrm{orb}$ can modify the morphology of the outflow resulting in a different angular-momentum loss than when the star is non-rotating.
This could potentially affect the evolution of the orbital parameters of the binary. 
On the other hand, whereas in Paper I we assumed a constant velocity profile of the wind, in this work the AGB wind feels an acceleration due to gas pressure, which results in a different wind velocity profile (see Paper II). 
However, we verified in a test that by taking a predefined terminal wind velocity, as in Paper I, this also results in an increase of the eccentricity.
Another difference compared to models Q2e0$i$ is that the radius of the donor is much smaller which could also impact the results.

A system with the parameters of model MMe05 could potentially counteract tidal circularisation in the region of interest for the progeny of AGB binary systems, since the orbital period is $\approx 1900$ days. However, as pointed out above, we should keep in mind that since the system is shrinking tidal interaction will become stronger.
In order to verify if the results of this model are physically possible a larger grid of simulations in which the stellar parameters of the donor star are computed with a stellar evolution code are necessary. 

\subsection{Other numerical and physical aspects}

\begin{table}
\centering
\caption{Numerical error in the angular momentum conservation}
\label{p3:table:dJ}
\begin{tabular}{c c}
\hline\hline
Model & $\delta J/J_\mathrm{init}$\\ 
- & yr$^{-1}$\\ \hline
Q2e0 & $8.62 \times 10^{-6}$ \\
Q2e02 & $8.79 \times 10^{-6}$ \\
Q2e04 & $8.94 \times 10^{-6}$  \\
Q2e06 & $9.74 \times 10^{-6}$\\
Q2e08 &  $4.96 \times 10^{-6}$ \\
Q2e06v1 &  $1.68 \times 10^{-5}$\\
MMe05 & $8.52 \times 10^{-6}$\\
\hline
\end{tabular}
\end{table}

Besides tidal interaction, there are other physical mechanisms and numerical aspects that could influence our results for the change in the orbital elements of the system. 
For instance, physical processes which have not been taken into account in this work are pulsations of the AGB star, dust formation and radiative transfer.
By considering these processes, the wind velocity profile will be different from that assumed in this work, which could result in a stronger interaction between the companion star and the wind, especially at periastron. 
This may affect the amount of angular momentum lost from the binary, and in consequence impact the evolution of the orbital parameters of the system. 
Furthermore, both pulsations and tides may induce a phase-dependence in the mass-loss rate, which can potentially lead to an increase of the orbital eccentricity \citep{Bonavic+2008}.

From the numerical point of view, we find that in some systems the large size of the sink particle results in an enhancement in the mass-accretion rate at different orbital phases (Sect.~\ref{p3:sec:only_beta}). 
One way to overcome this problem is by setting a smaller sink radius. 
However, this requires a correspondingly higher SPH resolution in order to prevent numerical noise. 
As seen in Paper I, a smaller sink radius will result in a smaller mass-accretion rate, but will not have a strong effect on the average angular-momentum loss. 
In the numerical simulations of wind mass transfer by \cite{shazrene_thesis}, the accretion process is modelled in a smooth fashion, which according to \cite{shazrene_thesis} provides a better numerical performance when the mass accretion rates are not constant as in the case of eccentric orbits. 
For instance, her numerical models of eccentric binaries show that, unlike most of our models in which the accretion disk is engulfed by the large sink, the accretion process occurs mainly when the stars are at periastron. However, in most cases she finds mass-accretion rates that are much larger than the BHL prediction (see discussion in Paper I). 

Another numerical aspect that may be of importance is the conservation of angular momentum. 
In Table \ref{p3:table:dJ} we show the error in the angular momentum budget for our calculations. 
Similar to Paper II, we find that angular momentum is not exactly conserved and that the error is larger for models in which strong interaction between the gas and the stars occurs. 
However, we stress that the errors shown in Table \ref{p3:table:dJ} correspond to the total angular momentum of the particles in the system (stars and gas), and that the error in the orbital angular momentum is only a fraction of this quantity, which unfortunately cannot be disentangled. 

We finish by noting that it is possible that not a single mechanism is responsible for the observed puzzling orbital periods and eccentricities of the descendants of AGB stars, such as Ba stars, CEMP-$s$ and post-AGB stars. 
In order to verify if an increase in eccentricity can occur in a regime with physically realistic parameters, a larger grid of numerical simulations with different binary masses, wind velocities, mass-loss rates, semi-major axes, and eccentricities is needed to reveal under which circumstances wind mass transfer can effectively counteract tidal circularisation and to understand under which circumstances other proposed eccentricity pumping mechanisms, such as the interaction with a circumbinary disk, may become important. 

\section{Summary}\label{p3:sec:conclude}

In this work we present the first exploratory hydrodynamical study of the impact of AGB wind mass transfer on the orbital parameters of eccentric low- and intermediate-mass binary systems. 
In order to do so we perform simulations using the \textsc{amuse} framework to couple a hydrodynamics code, which we use to model the wind dynamics, and an N-body code, which is used to model the dynamics of the stars. 
We explore a set of models with mass ratio $q=2$ and different combinations of semi-major axes and eccentricities, such that the donor star is close to filling its Roche lobe at periastron.

We find that for large eccentricities ($e\gtrsim 0.5$) the morphology of the outflow can be quite different from the circular case.
The spiral patterns found in the circular models or systems with small eccentricities become disrupted rings which move outward as the companion star makes its way through apastron. 
Furthermore, for large $e$ the outflow resembles the spherically symmetric wind case when the stars are near apastron. 

For models Q2e02 to Q2e08, where the initial wind velocity is $v_\mathrm{init} = 12$ km s$^{-1}$ and the eccentricity is varied between 0.2 and 0.8, we observe a similar trend in their orbital evolution in which $\dot{a}>0$ and $\dot{e} < 0$. 
On the other hand, in system Q2e06v1, with the same parameters as in model Q2e06 but $v_\mathrm{init} = 1$ km s$^{-1}$, we find that as periastron is approached a structure similar to wind Roche lobe overflow is formed. 
In this case the interaction between the wind and the companion star, as observed from the ouflow morphology, is stronger than in case Q2e06. 
In addition, we find that the average angular-momentum loss as well as the mass-accretion efficiency are higher than in model Q2e06 and both $\dot{a}<0$ and $\dot{e}<0$. 

Model MMe05, in which the stars are more massive and both the radius of the donor and the periastron distance are smaller, shows the most complex morphology among our models. 
In this case the orbit shrinks and the strongest interaction between the gas and the stars appears to occur near apastron, which results in an increase in eccentricity. 

Our results show a good agreement with the secular evolution equations derived by \cite{Dosopoulou+Kalogera2016} for fast isotropic winds, as long as the outflow in our models approximates the spherically symmetric wind case. 
However, when the ouflow morphology is modified by interaction with the binary, our results deviate from the analytical description. In these cases we find slower orbital expansion or faster orbital shrinking than predicted by the fast-wind approximation, and a slower decrease or even an increase in eccentricity.
We also find that the relation derived in Paper II for the angular-momentum loss as a function of the mass ratio and $v_\infty/v_\mathrm{orb}$ for the circular models agrees within $\approx 2-20\%$ when applied to the eccentric models by taking the average orbital velocity. The best agreement occurs when the wind velocity and the eccentricity of the system are large. 

Finally, we find that the hydrodynamical circularisation timescales are either longer than (for $e \lesssim 0.4$) or similar to the tidal circularisation timescales (for $e\gtrsim 0.6$). 
Given the short remaining time that a donor star such as assumed in our models spends in the AGB phase, a strong decrease in eccentricity will only occur for models with $e \lesssim 0.4$, whereas for models with larger eccentricity the change in the orbital parameters will be modest or almost negligible. 
Only for model MMe05 the hydrodynamical interaction could potentially counteract tidal circularisation. 
However, since the orbit in this model is shrinking, tidal effects may become stronger during the evolution of this system. 

A more detailed exploration of the parameter space of binary masses, wind velocities, semi-major axes and eccentricities is needed to find out under which circumstances wind mass transfer can effectively counteract tides and enhance the orbital eccentricity, and which role this process plays in explaining the puzzling orbital periods and eccentricities of the descendants of AGB binaries.

\begin{acknowledgements}
The authors thank the anonymous referee for helpful comments on the paper. 
MIS wants to thank Frank Verbunt, Chris Tout, Elliot Lynch, Avishai Gilkis and Glenn-Michael Oomen for the science discussions during the development of this project.
\end{acknowledgements}

\bibliography{bibliography}

\end{document}